\documentclass[pre,aps,twocolumn,
pdflatex,
superscriptaddress, floats, showpacs, 10pt]{revtex4-1}
\usepackage{graphicx}
\usepackage{dcolumn}
\usepackage{bm}
\usepackage{amssymb}
\usepackage{amsmath}
\usepackage{textcomp}
\usepackage{color}

\newcommand{\re}[1]{{\color{black}#1}}

\begin{document}

\title{High power gamma flare generation in multi-petawatt laser interaction with tailored targets}

\author{K.V. Lezhnin}
\affiliation{Department of Astrophysical Sciences, Princeton University, Princeton, New Jersey 08544, USA}
\affiliation{National Research Nuclear University MEPhI, Moscow 115409, Russia}

\author{P. Sasorov}
\affiliation{Keldysh Institute of Applied Mathematics RAS, Moscow 125047, Russia}
\affiliation{Institute of Physics ASCR, v.v.i. (FZU), ELI-Beamlines Project, Prague 182 21, Czech Republic}

\author{G. Korn}
\affiliation{Institute of Physics ASCR, v.v.i. (FZU), ELI-Beamlines Project, Prague 182 21, Czech Republic}

\author{S.V. Bulanov}
\affiliation{Institute of Physics ASCR, v.v.i. (FZU), ELI-Beamlines Project, Prague 182 21, Czech Republic}
\affiliation{National Institutes for Quantum and Radiological Sciences and Technology, KPSI, Kyoto 619-0215, Japan}
\affiliation{Prokhorov General Physics Institute, Russian Academy of Sciences, Moscow 119991, Russia}

\date{\today}

\begin{abstract}

Using quantum electrodynamics particle-in-cell simulations, we optimize the gamma flare ($\gamma$-flare) generation scheme 
from interaction of high power petawatt-class laser pulse with tailored cryogenic hydrogen target having extended preplasma corona. 
We show that it is possible to generate an energetic flare of photons with energies in the GeV range and total flare energy being on a kilojoule level 
with an efficient conversion of the laser pulse energy to $\gamma$-photons. 
We discuss how the target engineering and laser pulse parameters influence the $\gamma$-flare generation efficiency. 
This type of experimental setup for laser-based $\gamma$ source would be feasible for the upcoming high power laser facilities. Applications of high intensity $\gamma$ ray beams are also discussed.

\bigskip

\end{abstract}

\maketitle

\section{Introduction}
\label{INTRO}

In recent years, high power laser technology has reached the level of petawatt (PW) scale with kilojoule (kJ)  laser pulse energy \cite{PETAWATT, LFEX1, LFEX2}. 
Currently, ELI L4, a 10 PW, 1.5 kJ laser is being built at the ELI-Beamlines facility \cite{ELI-BL}, while the proposals for even higher laser pulse power facilities are announced. High power laser-matter interaction results in generation 
of high energy beams of charged particles, electrons and ions, photons in a wide frequency range spanning 
from low\re{-}frequency electromagnetic pulses to $\gamma$-rays. At high energy end, irradiation of plasma targets by high-intensity laser  leads to manifestation of nonlinear Thompson and Compton scattering processes, 
causing emission of photons with energies up to hundred MeV-scale, which is in $\gamma$ range. Generation of high power $\gamma$-flares is thought to be one of a primary goals for high power laser facilities,  \cite{ELI-BL, Gales2016, Turcu2016, GIST2018, SINGH}. Laser-based $\gamma$ ray source may be applicable in radiation chemistry and material sciences \cite{rad-chem1, rad-chem2},
in medicine in such a concept like 'gamma knife' \cite{GAMMAKNIFE1, GAMMAKNIFE2}, in nuclear physics, where $\gamma$-rays 
will help to excite isotopes \cite{ISOTOPE} for further use, as well as for laboratory astrophysics \cite{BULANOV2015} research, testing theories on astrophysical gamma 
ray bursts generation \cite{FIREBALL, PIRAN2005} and \re{behaviour} of a quantum electrodynamics (QED) plasma in pulsar magnetospheres \cite{GAMMAPULSAR}.

As theoretically foreseen, an irradiation of plasma targets by multi-petawatt laser radiation can result in high efficiency of the laser energy conversion to the energy 
of gamma\re{-}ray flash \cite{NAKAMURA2012, RIDGERS2012, SPIE2013, ZHU2016, 
VRANIC2016, GONG2017, HZLI2017, BOOST2018,Benedetti2018,MacchiPegoraro2018,Jansen2018}. Below, we present the results of multi-parametric studies of laser-target interaction for generation 
of bright $\gamma$-flare, aiming on parameters of $\approx 10$ PW, kJ-scale laser, which will be avail\re{a}ble in the coming years. 
Using quasi-classical \re{fully kinetic relativistic} 2D and 3D QED PIC 
simulations with the code EPOCH \cite{EPOCH}, we find a regime where a significant fraction of laser pulse energy may be 
converted to $\gamma$-rays by optimizing both preplasma and laser pulse. We show how target and laser parameters influence 
the $\gamma$ ray generation, specifying energy spectrum and angular distribution of low and high-energy photons. We provide 
analytical estimates for manifestation of Compton scattering processes in an underdense plasma medium and discuss its physics in detail.

The paper is organized as follows. In Section \ref{NTS-CO}, we reveal known properties of 
the nonlinear Thomson scattering and Compton scattering mechanisms of the ${\gamma}$ photon generation,
 which further will be used.
Then, in Section \ref{SETUP}, we formulate a numerical setup for our 2D and 3D QED PIC simulations. 
In Section \ref{RESULTS}, 
we discuss the simulation results and provide details of $\gamma$-flare optimization. In Section \ref{KINEMATICS}, analytical expressions for the inverse Compton scattering in a medium are derived. Finally, in Section \ref{CONCLUSIONS}, we restate our main 
findings and discuss further direction of the $\gamma$-flare generation research.

\section{Nonlinear Thomson Scattering and Compton Scattering Mechanisms of Gamma Photon Generation}
\label{NTS-CO}

In the case of tight  focusing of 10 PW laser pulses, the field intensity can reach values up to $10^{24}~ \rm W/cm^2$ corresponding to normalized field amplitude $a_0 = e E / m_e \omega_0 c \approx 10^3$, 
with $e$, $E$,  $m_e$, $\omega_0$, and $c$  being an elementary charge, electric field amplitude, electron mass, frequency of laser pulse, and speed of light in vacuum, respectively. 
This field amplitude is already enough for radiation reaction friction force to become dominant, as $a_0 \varepsilon_{\rm rad}^{1/3} >1$. 
Here the parameter $\varepsilon_{\rm rad}=4\pi r_e/3\lambda$ characterizes the radiation friction effects; $r_e=e^2/m_e c^2\approx 2.8 \times 10^{-13}\rm cm$ 
is the classical electron radius and $\lambda$ is the laser pulse wavelength (see \cite{QEDREVIEW, BULANOV2015} and references cited therein). 
The normalized field amplitude $a_0 =\varepsilon_{\rm rad}^{-1/3} $ for $\lambda=1\mu$m corresponds to the radiation intensity $\approx 10^{23}~ \rm W/cm^2$. 
The energy of photons emitted by ultrarelativistic electrons via the nonlinear Thomson scattering mechanism $\hbar \omega_{\gamma}$ is proportional to the cube of the electron energy,
\begin{equation}
\label{eq:NTS}
\hbar \omega_{\gamma}\approx 0.3 \hbar \omega_0 a_0^3.
\end{equation}
For $a_0\approx 200$ it is in $\gamma$-ray range. 

In the interval of laser amplitudes $1<a_0<\varepsilon_{\rm rad}^{-1/3}$
 the nonlinear Thomson scattering cross section grows as $\sigma_{\rm NTS}=\sigma_{\rm T}(1+a_0^2)$.
 Then at $a_0 \approx 1.1 \varepsilon_{\rm rad}^{-1/3} $ the radiation friction effects limit
 the cross section by the maximal value  $\sigma_{\rm NTS} =0.53 \sigma_{\rm T}\varepsilon_{\rm rad}^{-2/3} $. 
 For $ a_0>\varepsilon_{\rm rad}^{-1/3}$,
the cross section decreases as $\sigma_{\rm NTS}=\sigma_{\rm T}/a_0  \varepsilon_{\rm rad}$ (for details see Ref. \cite{BULANOV2015}).
Here, $\sigma_{\rm T}=(8\pi/3)r_e^2=6.65\times10^{-25}{\rm cm}^2$ is the Thomson scattering cross section. 

The gamma\re{-}rays, generated in laser plasmas 
due to the nonlinear Thomson scattering, were observed experimentally (see Refs. \cite{NTS0,NTS1, NTS2}).  
The \re{bremsstrahlung} mechanism can also generate the gamma-rays in this situation~\cite{GIZZI, GALY2007, G1P1, TEXAS2014, G1P2, WU}. However, the nonlinear Thomson and Compton effects are considerably more effective under conditions, discussed below.

When the energy of the photon \re{emitted} according to Eq. (\ref{eq:NTS}) becomes equal to the electron energy, the
 recoil effect cannot be neglected. Taking the \re{electron} energy to be equal to $m_ec^2 a_0$ we find from Eq. (\ref{eq:NTS}) that quantum regime regime starts at the laser amplitude above $\sqrt{m_ec^2/\hbar \omega_0}$, 
 i. e. at the intensity larger than $\approx 5\times10^{23}{\rm W/cm}^2$. The electron, colliding with the electromagnetic 
  wave, in this case, emits the gamma photons in the nonlinear or multi-photon Compton scattering regime.  
  The required intensity can be reached in the dense corona region, where the laser pulse undergoes the relativistic self-focusing, i.e. at later time of the laser-corona interaction. 

At an order of magnitude higher intensity, i. e. at  $10^{24}~ \rm W/cm^2$, when 
the dimensionless parameter $\chi_e \approx \gamma_{e,0} a_0/a_{\rm S}$ becomes larger than unity, $\chi_e>1$, 
such the QED effects as a recoil can play a significant role for a single electron interacting with the laser field.
Here $\gamma_{e,0}$ is the electron gamma factor and normalized Schwinger field, $a_{\rm S}=eE_{\rm S}/m_e\omega_0 c$ with $E_S=m_e^2c^3/e\hbar$, is $a_{\rm S}= m_e c^2/ \hbar \omega_0$.
The gamma photon radiation mechanism in this limit is the nonlinear or multi-photon Compton scattering.
 
The one-photon Compton scattering cross section is given by 
the Klein-Nishina formula~\cite{BLP}. In ultra-relativistic limit, when the parameter 
\begin{equation}
\label{eq:kappa}
\kappa=\left(\frac{\hbar \omega_0}{m_ec^2}+\gamma_{e,0} \right)^2
-\left(\frac{\hbar {\bf k}_0}{m_ec}+\frac{\hbar {\bf p}_0}{m_ec}\right)^2-1
\end{equation}
 is substantially larger than unity, total cross section is given by 
\begin{equation}
\label{eq:KN}
 \sigma_{\rm KN}=2\pi r_e^2 \frac{1}{\kappa}\left(\ln \kappa+\frac{1}{2}\right).
\end{equation}
Here $p_{||,0}$, $p_{\perp,0}$, and $\gamma_{e,0}=\sqrt{1+p_{||,0}^2+p_{\perp,0}^2}$ are the longitudinal and 
perpendicular, along and perpendicular to
electromagnetic wave propagation direction components of the electron momentum (${\bf p}_0=(p_{||,0},p_{\perp,0})$), 
and the electron gamma-factor before scattering, respectively. The photon frequency and wave-vector before 
 scattering equal $ \omega_0$ and ${\bf k}_0$. In nonrelativistic case, when $\kappa\ll 1$, i. e. the electron energy is 
  less than $30$~GeV, the Compton scattering cross section equals Thomson scattering cross section $\sigma_T$.

In the case of the Compton scattering on the electron in the field of the electromagnetic wave 
in vacuum, the dispersion equation for the wave frequency and wave vector takes the form $\omega^2={\bf k}^2 c^2$. 
In the ultra-relativistic limit, for $p_0\gg m_e c$ the parameter $\kappa$ given by Eq. (\ref{eq:kappa}) is approximately 
equal to $4 \hbar \omega_0 p_{||,0}/m_e^2 c^3$. 

According to Eq. (\ref{eq:NTS}) an electron interacting with the electromagnetic wave emits high order harmonics.
The maximum harmonic number could be equal to $a_0^3$. In quantum physics this corresponds to the electron
 interaction with $N_{ph}$ photons. Since an electron cannot emit the photon with the energy larger than the electron
  energy, $\hbar \omega_{\gamma} \leq m_ec^2 \gamma_e$, we obtain that the photon number is approximately 
  equal to $N_{ph}=(m_ec^2/\hbar \omega_0)a_0$. In this case, for $p_0\gg m_e c$ 
  the parameter $\kappa$ becomes equal to  $4 N_{ph} \hbar \omega_0 p_{||,0}/m_e^2 c^3=a_0 p_{||,0}/m_e c$.

The expression for the gamma 
ray photon energy can be found from the conservation of the energy and momentum in the photon-electron interaction:
\begin{equation}
\label{eq:Co-photon}
\hbar  \omega_{\gamma}=\frac{N_{ph}\hbar \omega_0(p_{||,0} c+ m_ec^2\gamma_{e,0})}
{N_{ph} \hbar \omega_0 + m_ec^2  \gamma_{e,0}- p_{\perp,0} c \sin{\theta} + (\hbar \omega_0 - p_{||,0}c) \cos{\theta}},
\end{equation}
i. e. in the expressions for one-photon Compton scattering $\hbar \omega_0$ should be replaced with 
$\hbar \omega_0=m_ec^2 a_0$.
The high energy $\gamma$-ray production in the multi-photon Compton scattering process 
has been observed in the experiments presented in Ref. \cite{BULA}. For theoretical aspects 
of  multi-photon Compton scattering see review articles \cite{QEDREVIEW, EHLOTZKY} and references therein. 
When the electron interacts with a plane electromagnetic wave of the amplitude $a_0$ we have 
$p_{\perp,0}=m_eca_0$ and $p_{||,0}=m_eca_0^2/2$ (e. g. see Ref. \cite{LL-TF}).

From Eq. (\ref{eq:Co-photon}) it follows that in the case of the head-on collision of the ultra-relativistic electron, 
$p_{||,0}\gg m_ec$, with the electromagnetic pulse the maximum emitted photon energy is equal to 
$N_{ph} \hbar \omega_0 4 (p_{||,0}/m_ec)^2\approx N_{ph}\hbar \omega_0 a_0^4$, 
provided $m_ec/2p_{||,0}=1/a_0^{-2}\ll N_{ph}\hbar \omega_0/m_e c^2$. 
It is equivalent to the condition $a_0\ll\sqrt{m_ec^2/ N_{ph}\hbar \omega_0}$. 
In the case $a_0\geq\sqrt{m_ec^2/N_{ph}\hbar \omega_0}$ the ${\gamma}$-photon energy 
is approximately equal to $\hbar  \omega_{\gamma}\approx p_{||,0}c$.
We see that in the limit $p_{||,0}/m_e c\gg 1$ the gamma photons are emitted at the angle $\theta_a$, 
which during a half of laser period changes from $\approx -  2/a_0$ to $\approx + 2/a_0$.
In the receding configuration, for co-propagating electron and laser pulse the photon energy is well \re{below} the energy 
of incident photon, $N_{ph}\hbar \omega_0/ 4 (p_{||,0}/m_ec)^2$. As discussed below in Section \ref{KINEMATICS} 
the plasma effects changing the \re{dispersion} equation provide the conditions for high energy photon generation in the co-propagating configuration too. 

Collective effects can make the threshold for the QED effects even lower, and, for laser pulses with $a_0 > 10^3$ interacting with plasma, in principle, 
 a self-consistent model of QED plasma should be developed \cite{MMPS, QEDREVIEW}.

\section{Simulation setup}
\label{SETUP}

\begin{figure}
 \includegraphics[width=1.05\linewidth]{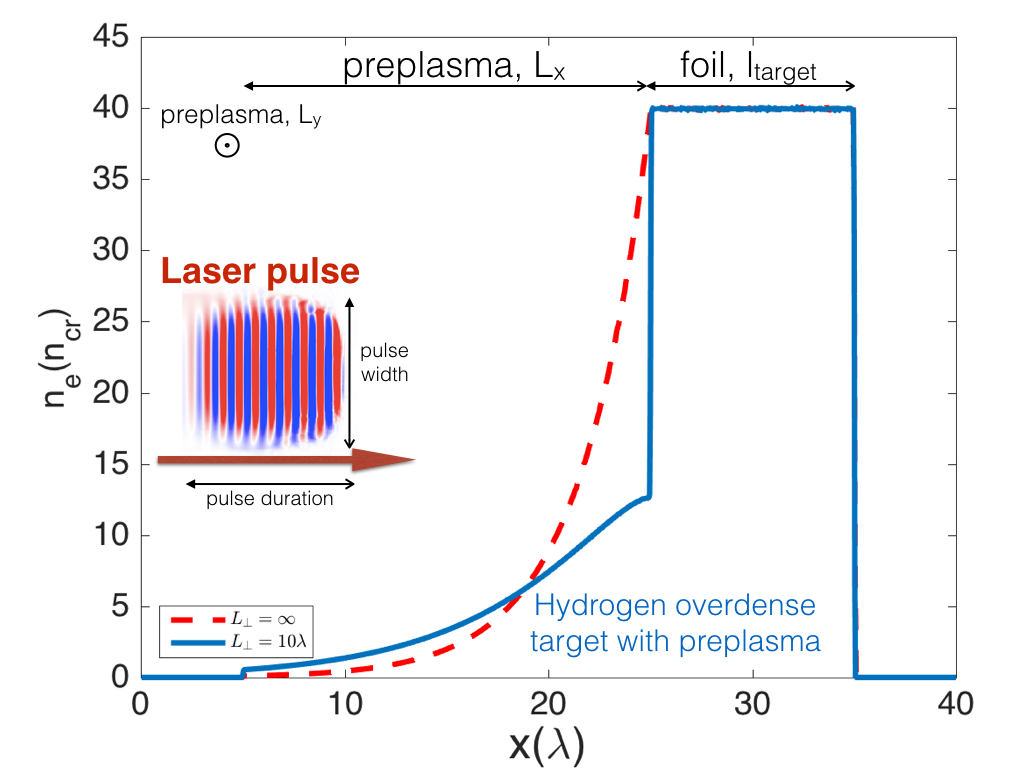}
\caption{Typical setup for 2D and 3D simulations. Target density profile averaged over transverse direction is shown for 
$L_\perp = \infty$ (red dashed line) and $L_\perp = 10 \lambda$ (blue solid line).}
\label{fig:init}
\end{figure}

The Particle-In-Cell (PIC) simulations are performed
with the relativistic electromagnetic code EPOCH \cite{EPOCH}, which includes QED processes. We perform a parametric scan 
with 2D version of the code, while also demonstrating results of the 3D simulation 
under the optimal conditions for $\gamma$-flare generation. 

In the 2D runs, we consider p- and s-polarized Gaussian pulse with peak laser pulse power in the range
from 1 to 20 PW. The pulse is incident on the target in the normal direction (it is along the $x$ axis), being focused
on the underdense corona \re{preceding} the high\re{-}density slab.  
We vary the pulse duration, from 5 fs to 150 fs, and the laser spot
size from $1 \lambda$ to $10 \lambda$. 
The optical axis of the laser pulse is at $y = 0$. 

The cryogenic hydrogen target \cite{TARGET} comprises of two parts. The overdense slab with the uniform density of $40 n_{\rm cr}$ and thickness of $l_{\rm target}$, varying from $1$ to $20 ~\lambda$ ($n_{\rm cr} = m_e \omega^2/ 4 \pi e^2$ 
 is the plasma critical density). The preplasma corona is localized at the front side of the foil.
 It has exponential distribution of the density, proportional to 
 $$\exp \left(-\left((x-x_0)^2/L_{\parallel}^2+(y-y_0)^2/L_{\perp}^2 \right) \right).$$ 
Here  ($x_0$, $y_0$) is a point on the front side surface of the high\re{-}density hydrogen slab 
and  $L_\parallel$ and $L_\perp$ are characteristic longitudinal and transverse scale-lengths of the corona density. 
 We cut the corona at $0.1 n_{\rm cr}$, and fix the length of the preplasma in the $x$ direction as 
 $L_{x} = L_{\parallel} \times \ln{(n_{\rm max}/n_{\rm min})}$. 
 In this expression, the maximum density equals $n_{\rm max} = 40~ n_{\rm cr}$ and the minimum
 density  $n_{\rm min}$ is chosen to be $0.1 n_{\rm cr}$. 
 We vary the transverse scale length $L_\perp$ from $1 \lambda$ to $\infty$. 
 Preplasma at the front side of the high\re{-}density slab is assumed to be formed by the ASE pedestal or/and by the prepulse, which alter an initial density distribution of the target for such a high laser pulse power and finite laser pulse contrast. 
 We choose the corona profile using the results of theoretical analysis of the preplasma corona formation presented in Ref. \cite{ESIRKEPOV2014}.
It is based on the hydrodynamics simulations conducted in order to describe the finite contrast effects 
 on the laser ion acceleration by petawatt laser pulse.
 Our simulation setup is also similar to the setup used in Ref. \cite{NAKAMURA2012}, where a 2D PIC simulations were conducted in order to show a feasibility 
 of the $\gamma$-flare generation. Our approach covers a broader range of parameters, and is based on a more fundamental 
 numerical model of QED processes \cite{QEDEPOCH}. 
 
 The \re{length} of a simulation box is $10\lambda + L_x + l_{\rm target}$, which varies from 
 $50 \lambda$ to $110 \lambda$. The transverse size of the box is $60 \lambda$. 
 We fix the number of grid points per $\lambda$ to be equal to 20, while doing some simulations with 40 grid nodes per $\lambda$. 
 We conduct a number of simulations varying the number of particles, which is typically $\{16,32\} \times 2.64\times 10^6$. \re{Total simulation time is $(10\lambda + L_x + l_{\rm target})/c + t_{\rm pulse}$, ranging from 200 to 500 fs. Here, $t_{\rm pulse}$ is the pulse duration.} \re{Total energy error is less than $5 \%$ in all simulations.} Initially, we don't have 
 any photons in our simulation, but they are being generated throughout the simulation via the nonlinear Thompson 
 and Compton scattering processes. The schematic of the simulation setup is shown in Figure \ref{fig:init}.

To find an optimal regime for $\gamma$-flare generation with the maximum efficiency of laser pulse 
energy conversion to the energy of the $\gamma$-photon we vary the simulation parameters as follows. 
The peak laser pulse power is within the interval (1,2.5, 5, 10, 20, and 40 PW). 
The pulse length equals 5, 10, 30, 50, and 150 fs. 
The laser pulse spot size, $\Delta w$ is equal to 10, 5, 2.5, and 1 $\lambda$. 
The laser pulse for 2D simulations has the s- and p- polarization. The thickness of the high 
density slab of the  target, $l_{\rm target}$ is equal to 1, 5, 10, and 20 $\lambda$.
The preplasma length, $L_x$, equals 1, 5, 10, 20, 40, 80, 160 $\lambda$, and the width, $L_\perp$,  
equals 1, 2.5, 5, 10, 20 $\lambda$ and $\infty$, i.e. in this case the plasma density 
is homogeneous in the transverse direction.

 We also conduct a series of auxiliary simulations in order to find whether or not 
 a strongly focused 10 PW laser pulse is able 
to generate a fair amount of the $\gamma$-photon energy during the interaction with solid density slab. This case corresponds to the scheme proposed in Ref. \cite{RIDGERS2012}.

In the case of 3D simulation, the parameters are as follows. 
We choose the simulation box size to be $70 \lambda \times 20 \lambda \times 20 \lambda$, with a grid 
resolution of 16 grid nodes per one $\lambda$. Total number of quasiparticles equals $1.84\times10^9$. 
The laser pulse peak power is of 10 PW. The laser \re{focal} spot size, 
$\Delta w$, is $2.5 \lambda$. The pulse duration is of 50 fs. The laser is incident on the target along the $x$ axis, centred around point $(y,z) = (0,0)$. 
The laser pulse is linearly polarized, with the electric field directed along the $y$ axis and the magnetic field along the $z$ axis. 
Hydrogen target is $5 \lambda$ thick, with preplasma size, $L_x$, equals to 10, 20, and 40 $\lambda$ and $L_\perp = \infty$. The simulation time equals 300 fs.

\begin{figure}
    \includegraphics[width=0.48\linewidth]{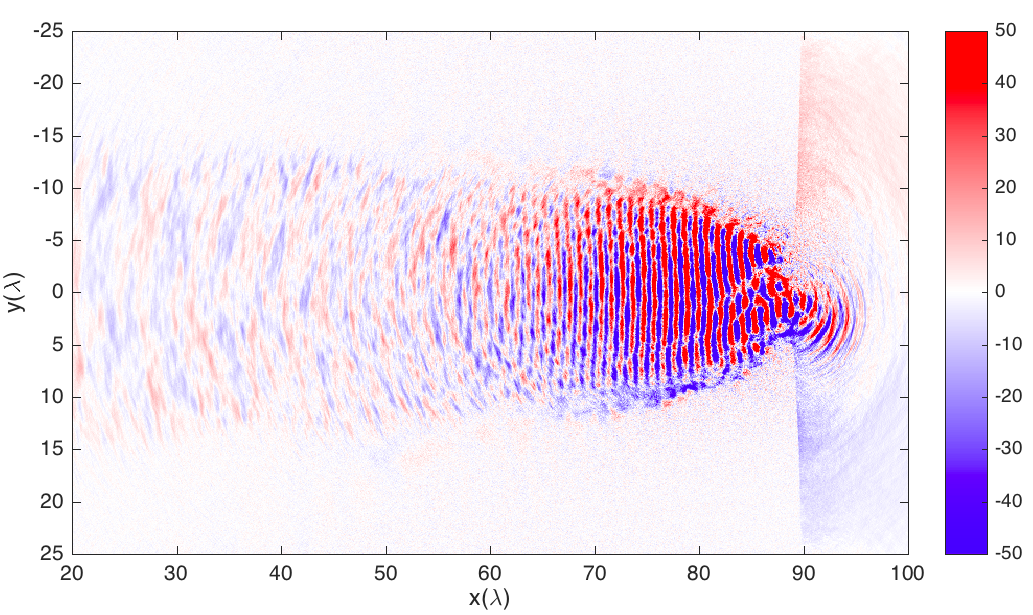}
    \includegraphics[width=0.48\linewidth]{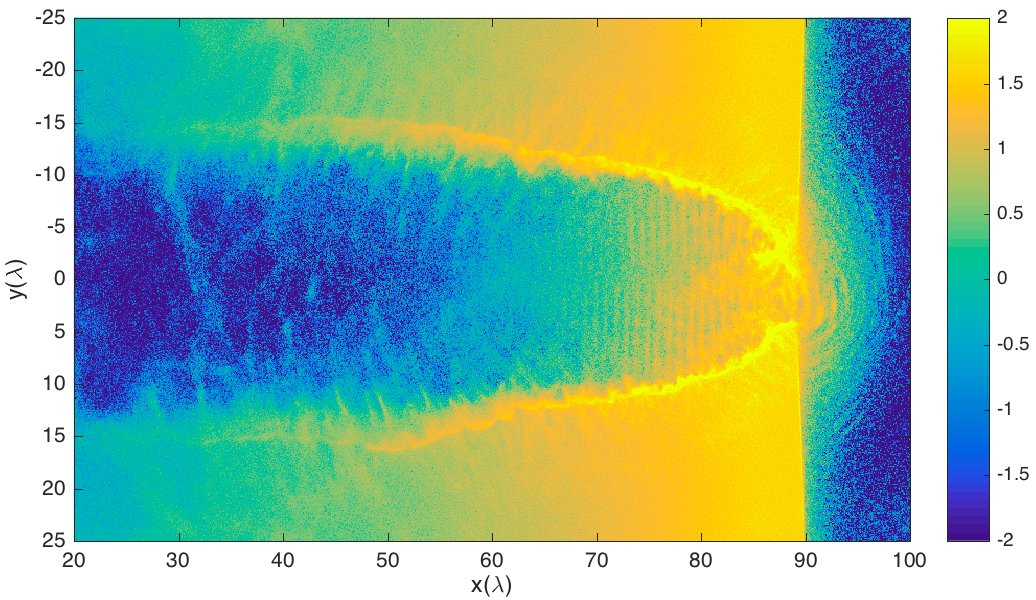}
    \par
    \includegraphics[width=0.48\linewidth]{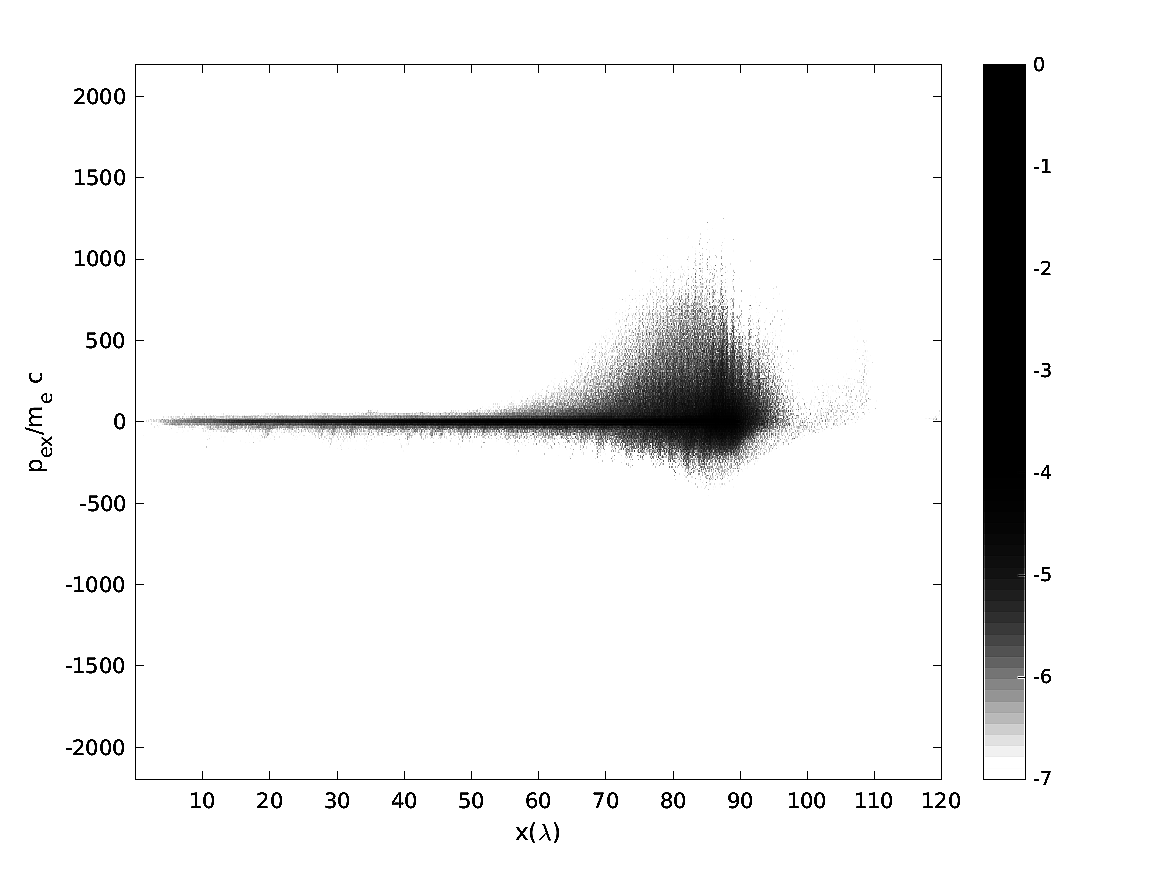}
    \includegraphics[width=0.48\linewidth]{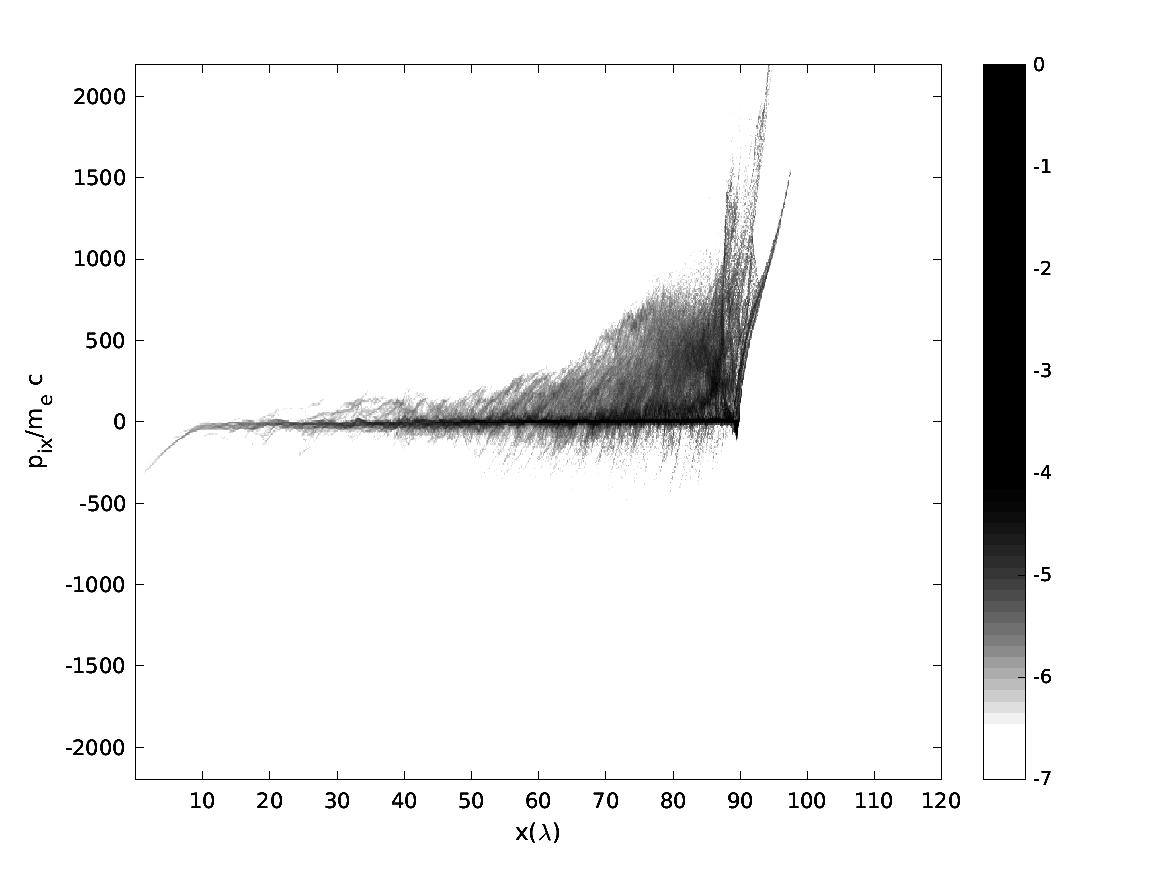}
    \par    
\caption{a) Distribution of the $z$- component of magnetic field, b) Distribution of $\rm  log_{\rm 10}$ of electron density, c) electron $x-p_x$ phase plane, d) ion $x-p_x$ phase plane for t = 400 fs. 
Pulse self-focusing, hole boring, electron heating, and ion acceleration via RPDA/TNSA mechanisms.}
\label{fig:plasma}
\end{figure}

\begin{figure*}
    \includegraphics[width=0.49\linewidth]{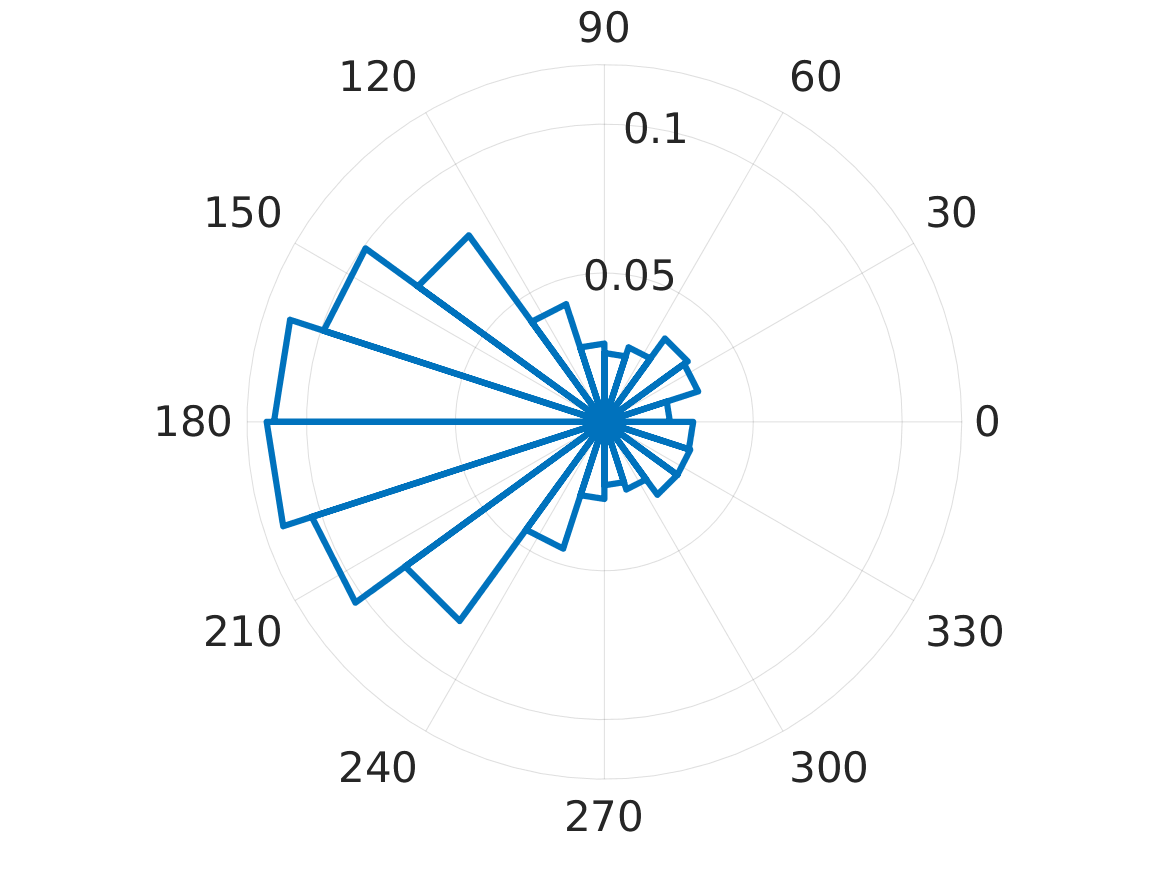}
    \includegraphics[width=0.49\linewidth]{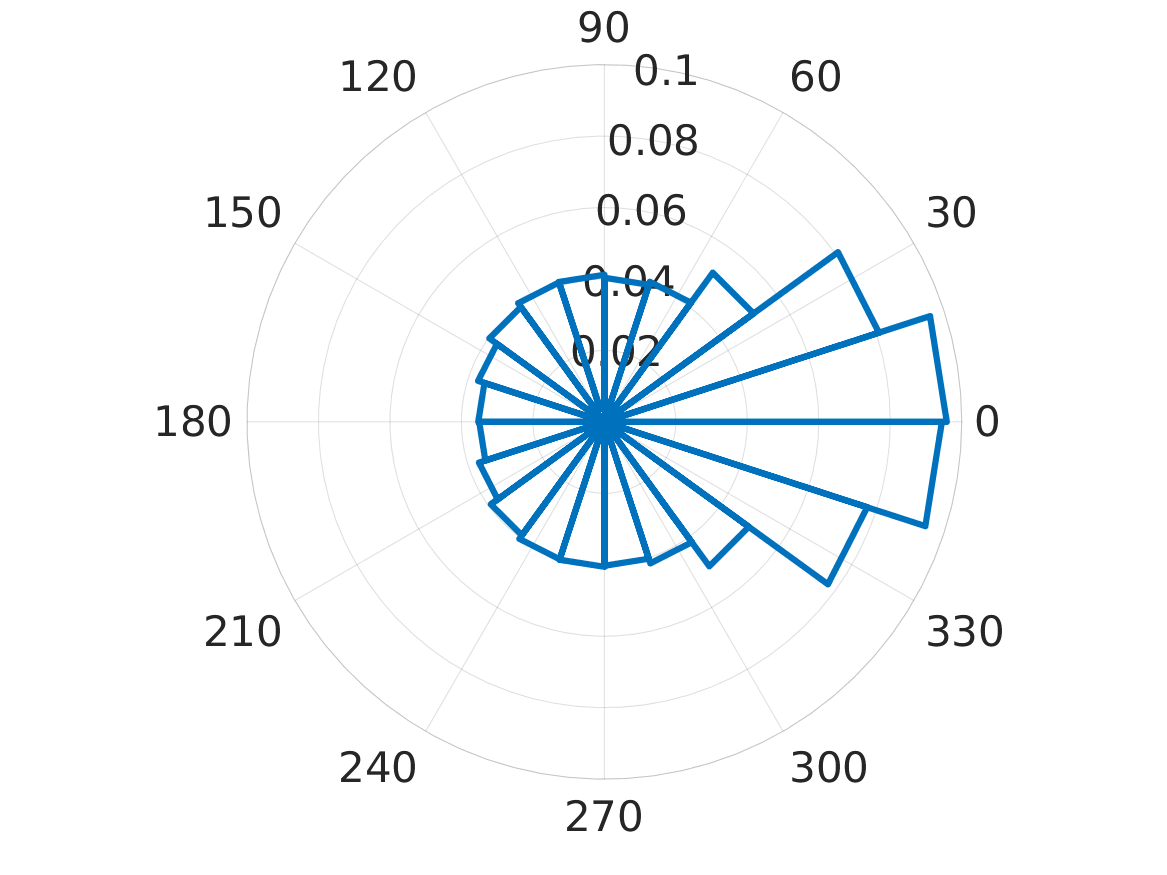}\par
    \includegraphics[width=0.49\linewidth]{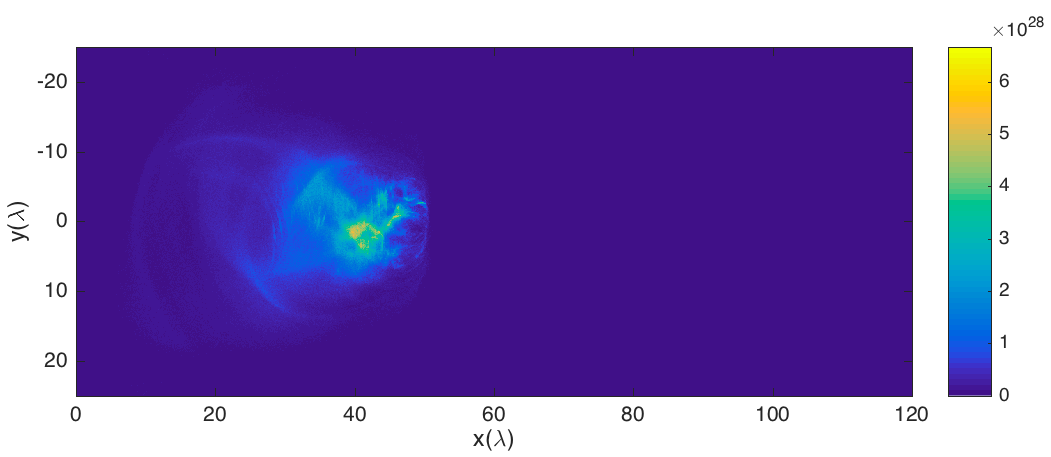}
    \includegraphics[width=0.49\linewidth]{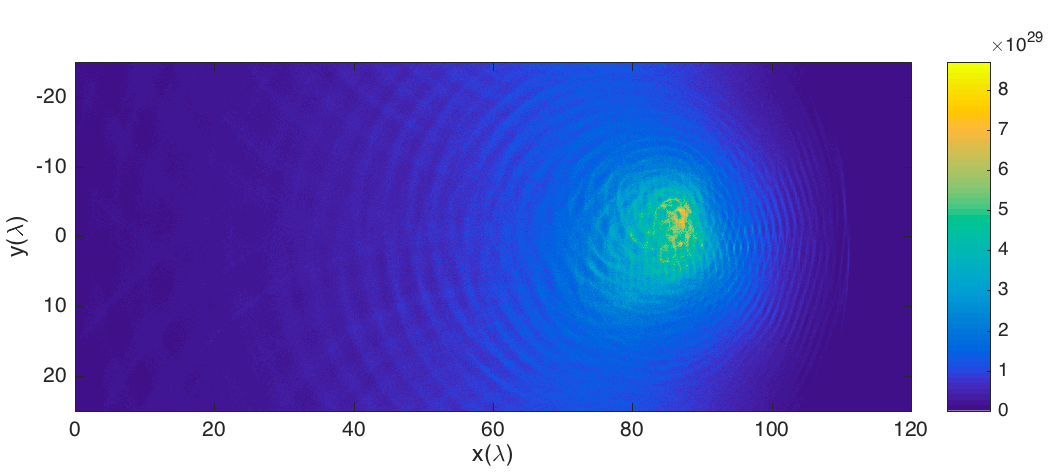}
\caption{Angular distribution of photon energy (a,b) and photon density distribution (c,d),  at t = 180 fs (a,c) and t = 400 fs (b,d). 
Angular {\it particle number} distribution is isotropic, thus, most energetic photons are directed backwards during the first stages 
of interaction, while the most energetic particles in the whole simulation are directed forward.}
\label{fig:gammaflare}
\end{figure*}

\begin{figure}
    \includegraphics[width=\linewidth]{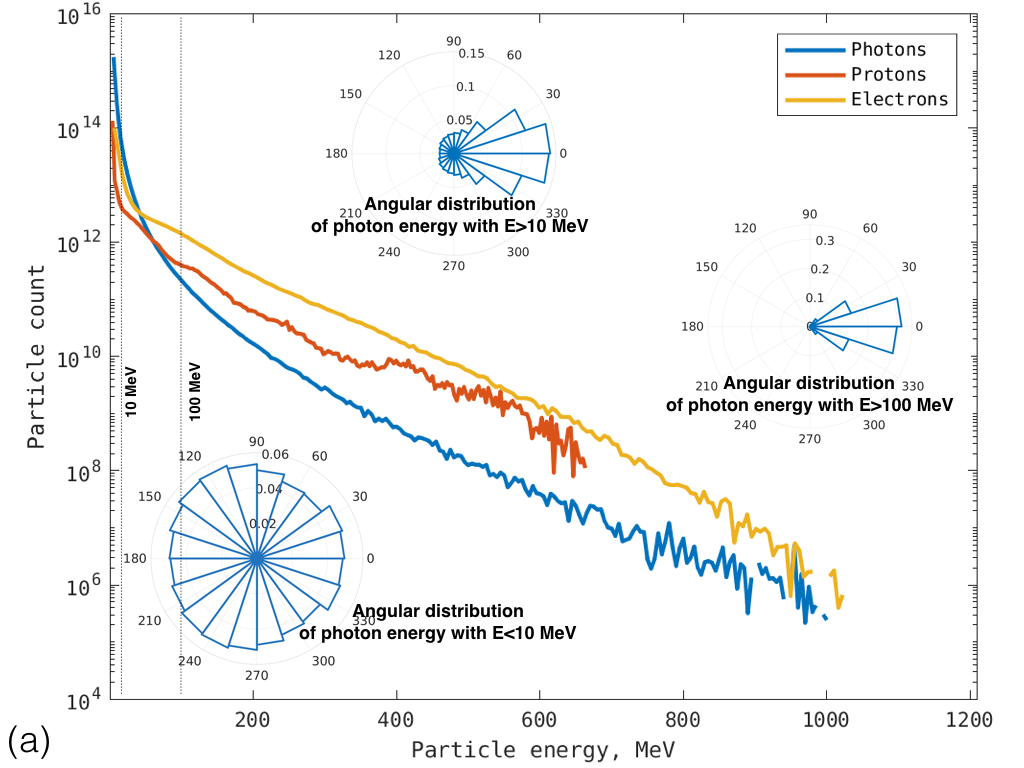}
    \includegraphics[width=\linewidth]{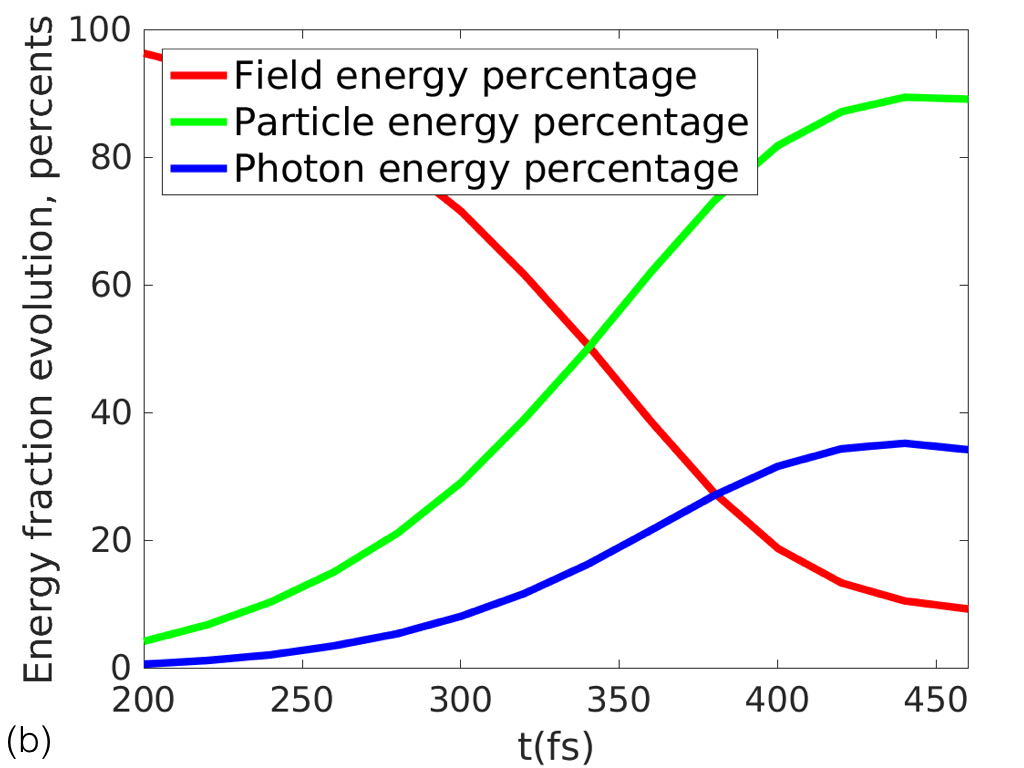}
\caption{a) Particle and photon energy spectra at t = 400 fs. Insets demonstrate an angular distribution of photon energy for photons with 
$\mathcal{E}_\gamma < 10 ~ \rm MeV$ (bottom left), photons with $\mathcal{E}_\gamma > \rm 10 ~ MeV$ (top center), 
and photons with $\mathcal{E}_\gamma > 100 ~ \rm MeV$ (middle right).  b) Particle, field, and photon energy evolution in the simulation.}
\label{fig:gammaflareenergy}
\end{figure}

\section{PIC simulation results}
\label{RESULTS}

Let us first discuss the 2D QED PIC simulation results.
Figure \ref{fig:plasma} shows a snapshot for $t=400 ~\rm fs$ from the simulation with the conditions optimal
in terms of the $\gamma$-flare generation. 
In this simulation run, the peak laser power is 10 PW, 
the laser pulse length equals 150 fs, the pulse width at the focus $\Delta w$ is equal to 2.5 $\lambda$. 
The radiation has p- polarization. The preplasma length in the longitudinal direction, $L_x$, is of 80 $\lambda$. 
In the transverse direction the plasma density is homogeneous, $L_\perp = \infty$.
The high\re{-}density slab thickness equals $5 \lambda$. 
As the width of the initial laser pulse is pretty small, $\lambda /\Delta w \approx 0.3$, the laser pulse diverges in transverse direction until 
it reaches a dense part of the preplasma, where it experiences the self-focusing.
As a result it reaches the high\re{-}density part of the target being focused, with the width equal to $2- 3 \lambda$, Figure \ref{fig:plasma} a. 
A normalized amplitude of the laser pulse at the front side surface may be as high as $a_0 \approx 300$. Laser pulse bores a hole in the target and partially \re{propagates} through it, as seen on Figure \ref{fig:plasma} b. 
Though the parameters \re{chosen} are not optimal for the ion acceleration, we still observe in $x-p_x$ phase plane 
\re{a} relatively high energy proton beam  (Figure \ref{fig:plasma} d). It can be formed as a result 
of the proton acceleration corresponding to Target Normal Sheath Acceleration (TNSA) \cite{TNSA} and/or the
Radiation Pressure Acceleration (RPA) \cite{RPDA} mechanisms. In the $x-p_x$ electron phase plane in 
Figure 2c, the electron heating up to 0.5 GeV energy in the $x$-direction is seen.  In opposite direction 
the electron energy is approximately equal to 125 MeV.
The maximum kinetic energy of protons is around 600 MeV.

 The angular distribution of photon energy is shown on Figure 3a(b) for $t=180~ (400) ~ \rm fs$. 
 Figure 3c(d) shows distribution of the  $\gamma$-photon density in the $x-y$ plane  at the instant of  time when the 
 maximum photon energy is reached:  $t=180 ~(400) ~ \rm fs$. 
 The main fraction of photon energy is shined in the laser pulse propagation direction. 
 It is worth noting that during a first few tens of femtoseconds, when the laser pulse propagates 
 in a very dilute plasma, the main fraction of the photon energy is directed {\it antiparallel} to the laser pulse propagation direction (Figure 3a). 
 These photons can be understood as produced by 
 scattering of laser pulse photons on a counterstreaming electrons, which try to circulate back along the self-focusing channel. Overall distribution of photon number may be considered almost isotropic, while
  the most energetic photons are emitted along the $x$ axis (Figure 3b).

Figure 4 shows the energy spectrum of particles at $t = 400~ \rm fs$ (Figure 4 a) and the energy evolution in a system throughout the simulation (Figure 4 b). 
We see that the preplasma allows us to almost completely absorb the laser pulse energy, leaving less than 10 \% 
in the electromagnetic field energy, which may partially be contained 
by the quasistatic magnetic field in the laser pulse wake. As is known the quasistatic 
magnetic field is associated with the electron vortices \cite{BULANOV1996}. In addition, not all the particle energy 
is converted into the $\gamma$-photon glow, as significant fraction is transformed into electron heating and ion acceleration (see Figures 2c, 2d, 4). 
However, for the parameters under discussion, the $\gamma$-flare optimization enables transforming $37 \%$ 
of the laser pulse energy into photons, with total ${\gamma}$ ray energy as high as $450~ \rm J$. 
Peak $\gamma$-flare power is around 3.8 PW. Comparison of photon and electron energy spectra shows that the emitted photon energies are approximately equal to electron energies. This fact  underlines the importance of a discrete, QED radiation reaction effects, in such a laser-plasma interaction setup. We have conducted an auxiliary set of simulations with laser pulse focused onto a spot of $1\lambda$ scale at the front side of the solid hydrogen slab,
 which is similar to the configuration used in Ref. \cite{RIDGERS2012}. We have concluded that
  the energy conversion efficiency is less than $20 \%$ in the case of a relatively thick target  with the thickness of $20 \lambda$.
  
  \section{Results of Multi-parametric simulations}

The results of multiparametric studies aimed at finding the maximal efficiency of the laser energy conversion  
to the $\gamma$ flash energy are presented in Figure 5. Here the laser plasma interaction parameters are 
changed in a broad range. The preplasma corona length is a key importance parameter for providing the multi petawatt laser pulse damping. As it follows from the simulations of the laser interaction with solid hydrogen target 
without a preplasma corona and for the target with a preplasma corona short compared to the laser pulse length, the laser radiation is mostly reflected back.
Typically approximately $50 \%$ of the laser is reflected, thus the $\gamma$ flash generation occurs under far from 
the optimal conditions. On the other hand, \re{a relatively long underdense corona in comparison} to laser pulse length is also not optimal for the $\gamma$ flash generation, since a large enough density gradient is required for \re{efficient} energy conversion. Auxiliary simulation with uniform density of $0.2 n_{\rm cr}$ shows a conversion efficiency
 less than $2 \%$ of laser pulse energy to photon energy. Apparently, there is an optimal preplasma length, 
 which is required for bright $\gamma$ radiation. The laser pulse length should also be long enough, as it is required for hole boring providing the conditions for high laser pulse to reach an optimal density gradient regions in the 
  preplasma corona. The laser pulse width is in tight connection with preplasma conditions, as the relativistic self-focusing for narrow laser pulses will eventually focus to a dense overcritical regions of the preplasma corona, where the laser-plasma interaction conditions are in agreement with estimates for an optimal laser power for efficient $\gamma$-flare, $\mathcal{P}_{\rm las} \approx 10^2 ~{\rm PW} \cdot n_{\rm cr}/n_e$, see \cite{BULANOV2010,NAKAMURA2012}. This condition yields that the 10 PW peak power of laser pulse should reach a overcritical density of $10 n_{\rm cr}$, which is indeed a time of peak power of $\gamma$-flare, according to Figure 4b.
\begin{figure}
    \includegraphics[width=1.05\linewidth]{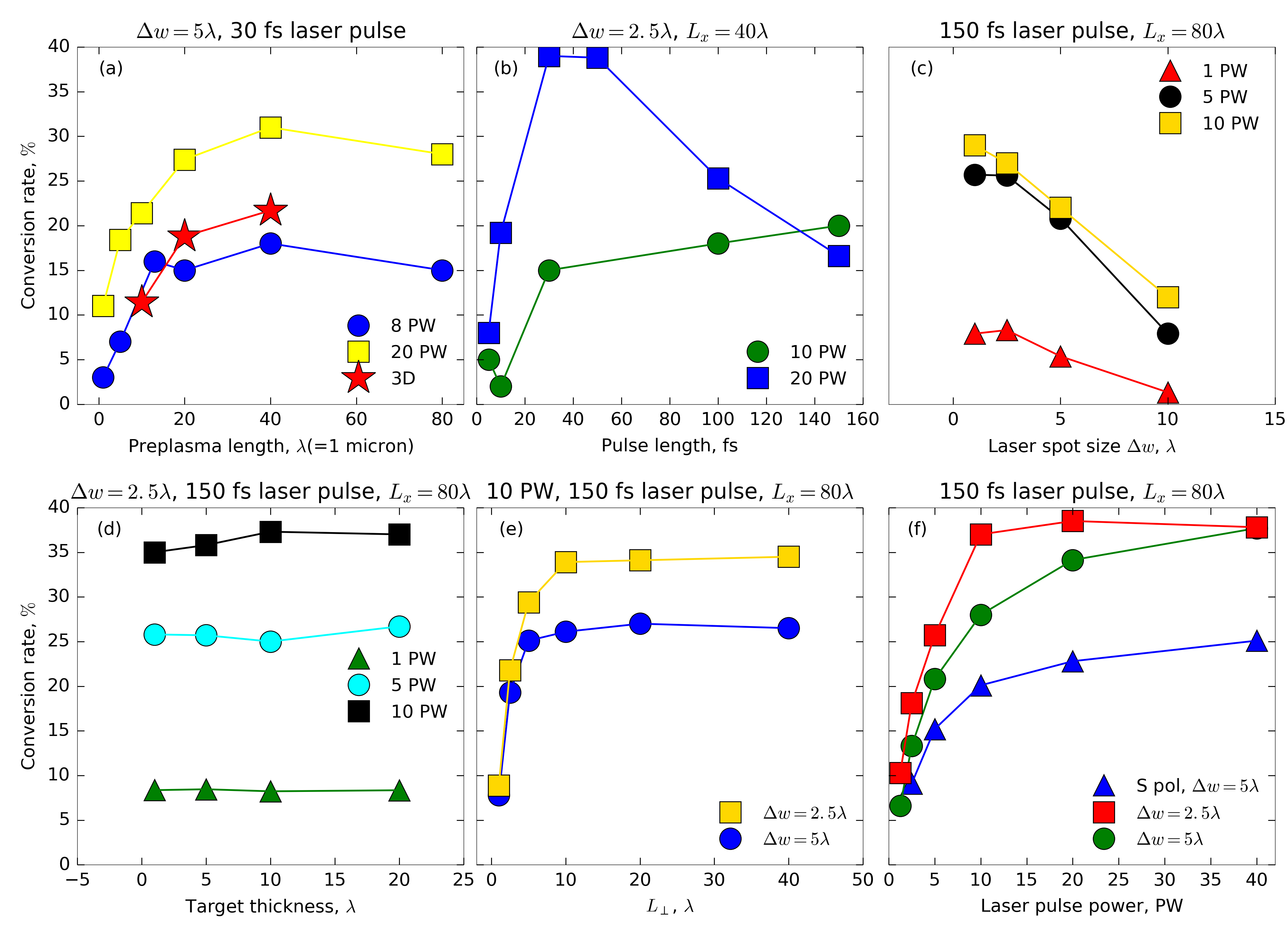}
    
    \caption{Optimization of laser pulse and hydrogen target parameters for efficient energy conversion to $\gamma$-rays. Each circle, triangle, star, and square corresponds to a specific choice of the PIC simulation parameters.
 Fixing all the parameters but one and varying it we plot the conversion efficiency dependence on this parameter. 
 The values of the parameters fixed in the simulation are written on each of the frames.
  Dependence of conversion efficiency on a) the preplasma length, in $\lambda$; 
  b) on laser pulse length for two peak laser pulse power values, 
  c) on the laser spot size for two peak laser pulse power values, 
  d) on the high\re{-}density slab thickness  for two peak laser pulse power values, 
  e) on $L_{\perp}$, which is an effective transverse width of the preplasma corona, for two laser spot size values, 
  f) on laser pulse power for two laser spot size values. }
\label{fig:gammaflareenergy1}
\end{figure}
\begin{figure*}
    \includegraphics[width=1.0\linewidth]{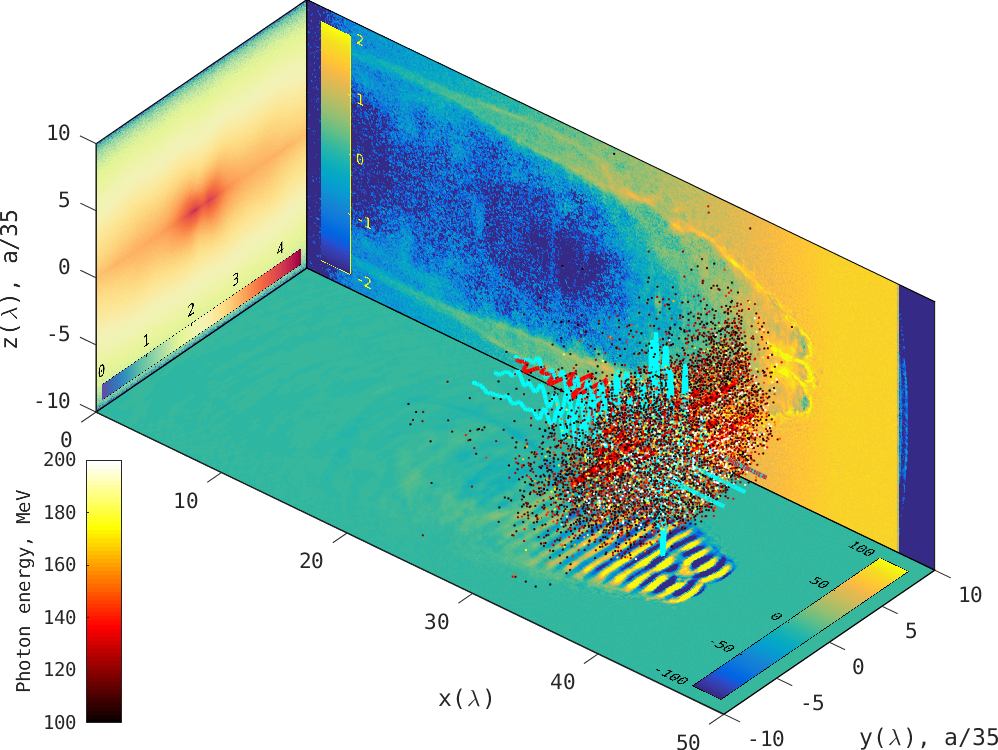}
    
    \caption{Snapshot of the 3D QED PIC simulation results for preplasma length of $L_x = 40 \lambda$, 
    with $\log_{10}$ of electron density distribution (in the $y=10 \lambda$ plane), the $z$ component of the magnetic field (in the $z=-10 \lambda$ plane), the angular distribution of $\log_{10}$ photon energy (in the $x=0$ plane), 
    and the high energy photon distribution shown with the points whose color corresponds to photon energy. 
    Cyan lines represent 1D cuts of the $z$ component of the magnetic field along the lines $y=0$, $z=0$; $y=-2, z=0$; $y=-4, z=0$. Red line represents a 1D cut of the $y$ component of the electric field along lines $y=0, z=0$. Colorbar shows a photon energy in MeV.}
\label{fig:gammaflareenergy2}
\end{figure*}

Among other parameters\re{,} the high\re{-}density target thickness provides only a minor contribution 
to manipulating the energy conversion efficiency, as by the time when the laser pulse starts to bore a hole 
in the high\re{-}density part of the target the most portion of its energy is already depleted, 
see Figure 2a and red line on Figure 4b. However, a larger target still allows one to contain more heated electrons within the high\re{-}density slab, leading to a slightly higher efficiency of the $\gamma$-flare generation.  
The transverse scale length of the preplasma, $L_\perp$, is determined by the spot size of the prepulse, as shown by radiation hydrodynamics simulations presented  in Ref. \cite{ESIRKEPOV2014}. Usually, $L_\perp$ is approximately
equal to the \re{focal} spot size of the ASE pedestal. It can limit the $\gamma$-flare generation efficiency, according to the parametric scan shown in Figure 5e. The laser pulse power is certainly \re{a} very important parameter, as it controls how efficiently laser pulse can propagate in preplasma, boring a hole there. Lower the power is, over less distance the pulse propagates through the preplasma, never reaching its corresponding optimum for $\gamma$-flare generation. 
For example, for $\mathcal{P}_{\rm las} = 2.5~ {\rm PW}$, the laser pulse should reach a $40 n_{\rm cr}$ region, but it is hard to do in the case of the target parameters used in the simulations.

Optimization of the preplasma profile for fixed 10 PW laser power allows one to obtain a $37 \%$ level 
of the laser energy conversion to the energy of $\gamma$-flare, which is slightly higher than it was obtained in previous 2D PIC simulations,  where a simple model 
of photon generation has been used \cite{NAKAMURA2012} with a simulation setup similar to the setup used in the present work. It is higher than the conversion efficiency found in previous 2D QED PIC simulations with EPOCH code \cite{RIDGERS2012}, but with different target setup.

Apparently the conditions used for the gamma\re{-}ray generation are not optimal for electron-positron pair production, as we do not consider laser pulse fields larger than $a_0=300$, 
while our additional simulations suggest that we need fields more than $a_{\rm pairs} \approx 10^3$ in order to see pair formation, while its density is still low to make significant influence 
on the $\gamma$-flare parameters, in agreement with Ref. \cite{RIDGERS2012}.

High-Z targets may be even more efficient in terms of \re{enhancement} of the gamma flare energy. As the simulations show\re{,} by keeping the same ion \re{number} density profile as in the optimal case, but for different target material, \re{such as copper or gold} with ionization degrees of 4 and 5, respectively, we can obtain an even higher energy conversion efficiency, up to 50\%. For such the targets, the reaching of a peak gamma flare power corresponds to the time when the laser pulse interacts with the overdense preplasma region whose density is approximately equal to
 $10 n_{\rm cr}$. A detailed studying of such type of the targets will be considered in a separate paper.

Finally, Figure 6 shows a snapshot of 3D simulation, which demonstrates $\gamma$-photons (colored circles with color corresponding the photon energy), 2D cut of $z$ component of the magnetic field through the $z=0$ plane, 
and  2D cut of electron density through the $y=0$ plane. In the $x=0$ plane, an angular distribution of photon energy, with the $y$ axis corresponding to $\theta$ angle changing from $-\pi$ to $\pi$ 
and $\phi$ angle, from $0$ to $\pi$. Cyan lines represent 1D cuts of the $z$-component of the magnetic field along the lines $y=0, z=0;~ y=-2, z=0;~ y=-4, z=0$. 
Red line represents 1D cut of the $y$ component of the electric field along the lines $y=0, z=0$.  We plot only high-energy photons with $\mathcal{E}_\gamma > 100 ~ \rm MeV$. Their primary location 
is around the region where the peak laser pulse power is. Here the low electron density cavity is located. Energy conversion \re{rate} is less than in an `optimal' case in 2D simulations, but still is \re{a significant fraction of laser pulse energy, being no less than $20 \%$.}
As in the case of 2D simulations, the majority of photon energy is directed forward along the laser pulse propagation direction. Photons are mainly confined within the cone with 60 degrees opening angle along the line $(y,z)=(0,0)$, where around $60\%$ of total photon energy is shined, see $x=0$ plane of Figure 6. The 3D simulations also show that longer corona allows to reach a better gamma flare generation efficiency.  

\section{Kinematics of inverse Compton scattering in collisionless plasma}
\label{KINEMATICS}

Gamma photons of a relatively low energy \re{(1-10 MeV and less)} are expected to be generated via the nonlinear Thomson scattering,
when the energy of emitted photons $\hbar \omega_{\gamma}$ is proportional to the cube of the electron energy (see Eq. (\ref{eq:NTS})).

Application of the theory of nonlinear Thomson scattering for high\re{-}efficiency gamma-ray generation has been 
discussed in details in Ref.  \cite{NAKAMURA2012}, where it was shown that the optimal laser power scales with the electron density in the plasma corona as $10^2(\omega_0/\omega_{pe})^2$~PW. In the case of the electron density
approximately equal to $10~ n_{cr}$, where the critical electron density is $n_{cr}=m_e\omega_0^2/4\pi e^2$, the 
required laser power is 10 PW.

As it is seen from Figures 2 and 4, at $t=400$ fs the gamma \re{-}ray angular distribution has a form of a collimated beam directed along the direction of accelerated electrons and laser light propagation. 
If the electrons were interacting in vacuum with co-propagating electromagnetic wave one could not expect significant generation of high energy photons. The situation changes in the medium with the 
refraction index corresponding to collisionless plasmas. In the medium the Compton scattering acquires new features \cite{Mackenroth2018}.

Here we consider a kinematics of the inverse multi-photon Compton scattering process in collisionless plasma of the near critical density when an ultra-relativistic electron collides with electromagnetic wave. 
By using the energy and momentum conservation in the electron-photon system we can find the scattering photon frequency dependence on the electron energy, the wave amplitude, the plasma density and the scattering angle.
The energy conservation equates the sum of electron and $N_{ph}$ photon energy before and after the scattering
\begin{equation}
\label{eq:energy}
m_ec^2\gamma+\hbar \omega_{\gamma}=m_ec^2\gamma_0+N_{ph} \hbar \omega_0.
\end{equation}
Here $\gamma_{e,0}=\sqrt{1+({\bf p}_0/m_ec)^2}$ and $\gamma_e=\sqrt{1+({\bf p}/m_ec)^2}$, 
${\bf p}_0$ and ${\bf p}$, and $\omega_0$ and $\omega_{\gamma}$ are the electron 
gamma factors, momenta, and photon frequency before and after collision, respectively.
The momentum conservation yields
\begin{equation}
\label{eq:momentum}
{\bf p}+\hbar {\bf k}_{\gamma}={\bf p}_0+N_{ph} \hbar {\bf k}_{0},
\end{equation}
where $\hbar {\bf k}_0$ and $\hbar {\bf k}_{\gamma}$ are the photon 
momentum before and after scattering on relativistic electron.

We assume that photon-electron interaction occurs in the $(x,y)$ plane, i. e.
${\bf p}_0=p_0 {\bf e}_x+m_ec a_0 {\bf e}_y$ and ${\bf p}=p_x {\bf e}_x+p_y {\bf e}_y$
with ${\bf e}_x$ and ${\bf e}_y$ being the unit vectors along the $x$- and $y$-axis, respectively, 
and $a_0=eE_0/m_e \omega_0 c$ normalized field amplitude of the laser radiation, which 
is assumed to be linearly polarized with the electric field parallel to the $y$-axis.

In collisionless plasmas the electromagnetic wave frequency and wave number are related to each other 
as $\omega=\sqrt{{\bf k}^2c^2+{\bar \omega_{pe}}^2}$, where 
${\bar \omega_{pe}}=\sqrt{4\pi n_e e^2/m_e\sqrt{1+a_0^2}}$ 
the plasma frequency ($n_e$ is the electron \re{number} density) with relativistic 
effects taken into account according to Ref. \cite{AkhPol}. 

Within the framework of the aforementioned assumptions by using Eqs. (\ref{eq:energy}) and (\ref{eq:momentum}) 
and assuming that $\omega_0=\bar \omega_{pe}$, i.e. the electron interacts with the electromagnetic wave in 
the critical density region, we find  that the scattering cross section equals to that given by Eq. (\ref{eq:KN})
with the parameter $\kappa$ defined by Eq. (\ref{eq:kappa}) approximately 
equal to to $2 N_{ph} \hbar \omega_0 p_{||,0}/m_e^2 c^3$. 

\begin{figure}
    \includegraphics[width=1.05\linewidth]{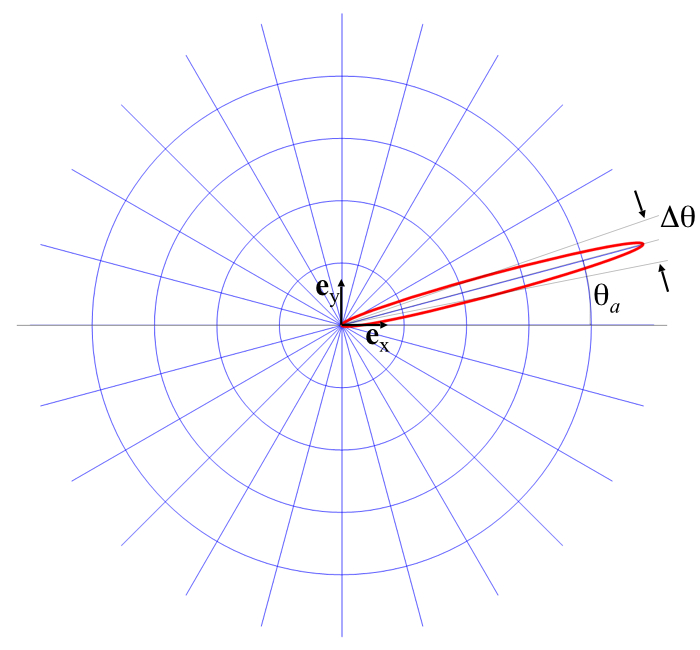}    
    \caption{\re{Angular} distribution of gamma-photons emitted at the angle $\theta_a$ and confined withing the cone of the angle $\Delta \theta$. 
    \label{fig:omegagamma}}
\end{figure}

The energy of the gamma-photon generated 
in the Compton scattering process is given by
\begin{equation}
\label{eq:photon-g}
\hbar \omega_{\gamma}=\frac{N_{ph} \hbar \omega_0(\hbar \omega_0+ m_ec^2\gamma_{e,0})}
{N_{ph} \hbar \omega_0 + m_ec^2 \gamma_{e,0}- p_{\perp,0} c\sin{\theta}  - p_{||,0}c \cos{\theta} }.
\end{equation}
Here $\theta$ is a scattering angle, i.e. 
$ {\bf k}_{\gamma}=|{\bf k}_{\gamma}|( {\bf e}_x \cos\theta + {\bf e}_y \sin\theta)$. 
 We see from Eq.~(\ref{eq:photon-g}) that in the limit $p_{||,0}/m_e c\gg 1$ and $p_{\perp,0}=m_ec a_0$ the gamma photons are emitted at the angle $\theta_a$, 
which during a half of laser period it changes from $\approx - a_0/p_{||,0}$ to $\approx + a_0/p_{||,0}$ being confined within the cone 
$\Delta \theta\approx \sqrt{ (1+a_0^2) m_e^2 c^2/p_{||,0}^2+N_{ph} \hbar \omega_0/m_e c^2}$. The \re{angular} 
dependence of the ${\gamma}$ photons, when their distribution is alongated according to the 
laser pulse polarization has been observed in the experiment presented in Ref. \cite{NTS2}. 
It is also well seen in Fig.~\ref{fig:omegagamma}.

The \re{maximum} gamma photon energy is approximately equal to $N_{ph} \hbar \omega_0 2 (p_{||,0}/m_ec)^2$ provided $p_{||,0}/m_ec\ll m_ec^2/2N_{ph}\hbar \omega_0$. Otherwise, for
$p_{||,0}/m_ec\gg m_ec^2/2 N_{ph}\hbar \omega_0$ the gamma photon energy is about 
$\hbar  \omega_{\gamma}\approx m_e c^2 \gamma_{e,0}$, which is a typical case for the parameters under the consideration \re{ - these photons may obtain energies up to GeV level.}

\section{CONCLUSIONS AND DISCUSSIONS}
\label{CONCLUSIONS}

In this paper, the optimization of laser-plasma interaction parameters for efficient $\gamma$-flare generation, in terms of both laser pulse energy conversion to 
$\gamma$-rays and $\gamma$-flare power are discussed. The multiparametric analysis based on the using 2D and 3D QED PIC code EPOCH shows how the laser pulse and preplasma parameters are related to the $\gamma$ radiation from the laser irradiated target. 
Typical target under consideration \re{comprises} the solid density hydrogen slab with the near-critical density inhomogeneous preplasma corona. 
The corona length is approximately equal to the laser pulse length, i. e. it 
is typically in the interval from $10\mu$m to $50\mu$m. For such the corona plasma parameters the bremsstrahlung 
energy losses of ultrarelativistic electrons are negligibly week. The electrons \re{lose} the energy here in the form 
of the high\re{-}energy photons. The photon radiation mechanism is the nonlinear Thomson scattering, and at the 
electron high energy end the $\gamma$-photons are emitted via the multi-photon Compton scattering with the 
photon number $N_{ph}\gg 1$ . 

The simulations provide an information on the angular distribution of photon energies in different energy bands. 
It is shown that the low-energy $\gamma$-photons with energies less than $10 ~ \rm MeV$ 
are \re{distributed} isotropically, while the high energy photons are directed mainly in the direction of laser pulse propagation. 
Maximum energy of photons may reach a GeV energy level. In the process of high energy 
photon generation in the co-propagation electron-electromagnetic wave configuration the crucial role is played 
by the fact that the group velocity of the laser pulse in the near-critical density region becomes substantially less
than the speed of light in vacuum. As we may see from Eq. (\ref{eq:photon-g}) in this case the high energy photons 
can be generated in both the electron-photon co-propagation and counter-propagation configurations. 
This is in contrast with the electron-photon interaction described by Eq. (\ref{eq:Co-photon}).

\re{Regarding} the ion acceleration for the laser-target parameters under discussion, it occurs \re{to be in a non-optimal} regime,
 as we seek for conditions to maximize the energy conversion to $\gamma$-photons. 
However, the ion acceleration up to 600~MeV is seen via a combination of the RPA and TNSA acceleration mechanisms. 
Electron heating is also seen, up to GeV level. Analyzing the electron and photon energy spectra, we may conclude
that with the 10 PW pulse laser\re{, the QED radiation reaction regime can be reached, where radiation reaction 
affects the electron dynamics by emitting discrete photons with energies up to the electron energy, rather than its classical version of continuous radiation. }

The 3D simulations prove the conceptual feasibility to generate a bright $\gamma$-flare, with energy conversion 
rate of around $20 \%$, which is in agreement with our 2D simulations with corresponding laser pulse and target parameters. 

Our findings open a way towards various applications of ultra-short pulse high power $\gamma$ ray sources. 
Among them one of the most attractive is the material sciences allowing one to extend the radiation 
chemistry \cite{rad-chem1, rad-chem2} to the regimes when ultra-relativistic physical processes come into play.
In the GeV gamma\re{-}ray energy range the gamma\re{-}rays are absorbed in the media via the electron-positron pair
generation \cite{gamma-CS} with the cross section approximately equal to 10 barns. In contrast with other $\gamma$
ray sources, in the case of the laser generated $\gamma$-rays, the flash duration is in the \re{femtosecond to picosecond} interval.

\section{Acknowledgements}

Computational resources were provided by the ECLIPSE cluster of ELI-Beamlines. 
The EPOCH code was developed as part of the UK EPSRC funded projects EP/G054940/1. 
At ELI-BL, the work has been supported by the 
project High Field Initiative
(CZ.$02.1.01/0.0/0.0/15\_003/0000449$) from the European
Regional Development Fund.
KVL is grateful to ELI-Beamlines project for hospitality during this work. 
Authors are thankful to T. Zh. Esirkepov, V. Istokskaia, 
J. Koga, K. Kondo, M. Kando, T. Kawachi, D. Margarone,
K. Nishihara, F. Schillaci, S. Singh, K. Tanaka, and W. Yan
for useful discussions and to
 Y. J. Gu, D. R. Khikhlukha, E. Chacon-Golcher, 
and M. Matys for assistance with computer simulations.

\end{document}